\documentclass[12pt,twoside]{article}
\usepackage{amssymb}
\def\1#1{{\bf #1}}
\def\2#1{{\cal #1}}\def\3#1{{\sl #1}}\def\4#1{{\tt #1}}\def\5#1{{\sf #1}}
\def\6#1{{\mathfrak #1}}\def\7#1{{\mathbb #1}}

\def\aut{\rm Aut}

\def\ska{\vskip 0.2cm}

\def\beq{\begin{equation}}
\def\eeq{\end{equation}}
\def\vs{\vspace{0.2cm} \\}
\newtheorem{Def}{Definiton}[section]
\newtheorem{Lem}{Lemma}[section]
\newtheorem{The}{Theorem}[section]
\newtheorem{Pro}{Proposition}[section]
\newtheorem{Cor}{Corollary}[section]
\def\bdef{\begin{Def}\1: \em}
\def\eef{\end{Def}\ska}
\def\blem{\begin{Lem}\1: }
\def\elem{\end{Lem}\ska}
\def\bthe{\begin{The}\1: }
\def\ethe{\end{The}\ska}
\def\bpro{\begin{Pro}\1: }
\def\epro{\end{Pro}\ska}
\def\bcor{\begin{Cor}\1: }
\def\ecor{\end{Cor}\ska}
\def\supp{{\rm supp}}
\def\sp{{\rm  sp}}

\def\id{{\sf id}}

\def\tr{{\sf tr}}
\def\al{\alpha}

\def\gam{\gamma}\def\Gam{\Gamma}

\def\te{\theta}

\def\sgm{\sigma}
\def\om{\omega}\def\Om{\Omega}

\def\ra{\Rightarrow}

\def\bpr{\noindent{\em Proof. }}

\def\epr{$\square$\ska}

\def\tl{\tilde}

\def\<{\langle}
\def\>{\rangle}
\def\Ad{{\sf Ad}}

\begin{document}
\title{\bf On the Existence of Kink-(Soliton-)States}
\author{\it Dirk Schlingemann \\ II.Institut f\"ur Theoretische Physik \\
Universit\"at Hamburg
\\ \vs\vs 
DESY 95-239  \\
hep-th/9512100}
\date{}
\maketitle
\abstract{
There are several two dimensional quantum field theory models 
which are equipped with different vacuum states. For example
the Sine-Gordon- and the $\phi^4_2$-model.  
It is known that in these models there are also 
states, called soliton- or kink-states, 
which interpolate different vacua. We consider the following question:
Which are the properties a pair of vacuum sates must have, such that an 
interpolating kink-state can be constructed?
Since we are interested in structural aspects and not in specific details 
of a given model,
we are going to discuss this question in the framework of algebraic 
quantum field theory which includes, for example, the $P(\phi)_2$-models. 
We have shown that for a large class of vacuum states, including the vacua of 
the $P(\phi)_2$-models, there is a natural way to 
construct an interpolating kink-state.}

\section*{Introduction}
In quantum field theory there are several models in $1+1$ dimensional 
space-time
which are equipped with different vacuum states. For example
the Sine-Gordon-model, the $\phi^4_2$-theory and the Skyrme-model in 
$1+1$ dimensions. Further candidates are special types of $P(\phi)_2$-models.
It is known that in the Sine-Gordon- and  $\phi^4_2$-model there are also 
states which interpolates different vacuum states. These states are called
soliton- or kink-states. An construction in the framework of 
algebraic quantum field theory was done by J. Fr\"ohlich \cite{Froh1} in the 
70s where he also discussed the existence of kink-states in general
$P(\phi)_2$-models. However, these construction leads only to 
kink-states which interpolates vacua which are connected 
by an (special) internal symmetry transformation. Since there are also 
candidates for models with different vacua which can not be connected by 
a symmetry transformation, we can ask the following question:

If we consider any quantum field theory in $1+1$ dimensions which is 
equipped with different vacuum states. 
Which are the properties a pair of vacuum sates must have, such that an 
interpolating kink-state can be constructed?

Since we are interested in structural aspects and not in specific details 
of a given model,
we are going to discuss this question in the framework of algebraic 
quantum field theory which includes, for example, the $P(\phi)_2$-models. 
For this purpose, let us describe the main aspects of algebraic quantum 
field theory in $1+1$ space-time dimensions. 

A $1+1$ dimensional quantum field theory 
is given by a prescription which assigns to each region $\2O\subset\7R^2$
a C*-algebra $\6A(\2O)$ and the elements in $\6A(\2O)$ represent 
physical operations which are localized in $\2O$. 
These prescription has to satisfy a list of axioms
which are motivated by physical principles.
\begin{description}
\item[{\it (1)}]
A physical operation
which is localized in a region $\2O$ should also localized in each 
region which contains $\2O$. Therefore, we require that   
if a region 
$\2O_1$ is contained in a lager region $\2O$, then the algebra $\6A(\2O_1)$
is a sub-algebra of $\6A(\2O)$. 
\item[{\it (2)}]
Two local operations which take place 
in space-like separated regions should not influence each other.
Hence the {\em principle of locality} is formulated as follows:
If a region 
$\2O_1$ is space-like separated from a region $\2O$, 
then the elements of $\6A(\2O_1)$ commute with those of $\6A(\2O)$. 
\item[{\it (3)}]
Each operation which is localized in $\2O$ should have an 
equivalent counterpart which is localized in a translated region
$\2O+x$. The {\em principle of translation covariance} is described by 
the existence of a 
two-parameter automorphism group $\{\al_x;x\in\7R^2\}$ which 
acts on the C*-algebra $\6A$, 
generated by all local algebras $\6A(\2O)$, such that 
$\al_x$ maps $\6A(\2O)$ onto $\6A(\2O+x)$.
\end{description}
A prescription $\2O\to\6A(\2O)$ of this type is called a {\em translationally
covariant Haag-Kastler-net}.
To introduce the notion of kink-states, 
we first should discuss the notion of a (physical) state in our framework. 
A state is a positive linear 
functional $\om$ on $\6A$ with $\om(1)=1$.
Using the {\em GNS-construction}, we obtain a Hilbert-space $\6H$, a
representation $\pi$ of $\6A$ on $\6H$ and a vector $\Om$, such that 
$\om(a)=\<\Om,\pi(a)\Om\>$ for each $a\in\6A$. There may be many 
states on $\6A$ and we need a criterion to select the states of 
physical interest. For our purpose, we use the {\em Borchers-criterion}
which requires:
\begin{description}
\item[{\it (1)}] There exists a unitary strongly continuous representation
of the translation group $U:x\mapsto U(x)$ on the GNS-Hilbert-space $\6H$ 
which implements $\al_x$ in the GNS-representation $\pi$, i.e.
$\pi(\al_xa)=U(x)\pi(a)U(-x)$, for each $a\in\6A$.
\item[{\it (2)}] The spectrum (of the generator) of $U(x)$ is contained 
in the closed forward light cone.
\end{description}

A special class of states which satisfy the Borchers-criterion are
the {\em vacuum-states} which fulfill the additional property of 
translation invariance, i.e. $\om\circ\al_x=\om$.

We are now prepared to describe the properties of a kink-state:
\begin{description}
\item[{\it Particle-like properties:}] 
We require that a kink-state fulfills the Borchers-criterion. These property
guarantees that one has the possibility to "move" a kink like a particle.
If the lower bound of the spectrum of $U(x)$ is 
an isolated mass-shell, then a kink-state "behaves" completely like 
a particle. 
\item[{\it The interpolation property:}]
A pair of vacuum states $\om_1,\om_2$ is interpolated by a 
kink-state $\om$ if there is a bounded region $\2O\subset \7R^2$, 
such that $\om(a)=\om_1(a)$, if $a$ is localized in the 
left space-like complement of $\2O$, and $\om(a)=\om_2(a)$,
if $a$ is localized in the right space-like complement of $\2O$. 
In other words, in one space-like direction $\om$ "looks like" the 
vacuum $\om_1$, in the other space-like direction $\om$ "looks like" 
the vacuum $\om_2$.
\end{description}

We come now to the question which conditions a pair of vacuum-states
has to satisfy, such that a interpolating kink-state can be constructed.

In the first two sections we introduce the notion of 
{\em admissible, locally equivalent} 
vacuum-states. 
As we will see
later, these condition is   
sufficient for the existence of an interpolating
kink-state. 
As an example of an {\em admissible} vacuum-state, one 
may think of the vacuum of a massive free scalar 
field \cite{Bu1,AntFre,Schl3}. But for a direct application of our 
construction to the $P(\phi)_2$-models, one has to check that these
property remains also true for the interacting case.

Therefore, we consider also a lager class of  
vacuum-states which we call {\em weakly admissible}.
The disadvantage of these condition is that it is much more technical.
On the other hand, it 
can be proven for the vacuum-states of $P(\phi)_2$-models \cite{Schl4}.

The main result of this paper is presented in section 3. It states that 
for each pair of {\em weakly admissible} 
vacuum states there exists a kink-state 
which interpolates them. 
The following diagram shows the logical structure:
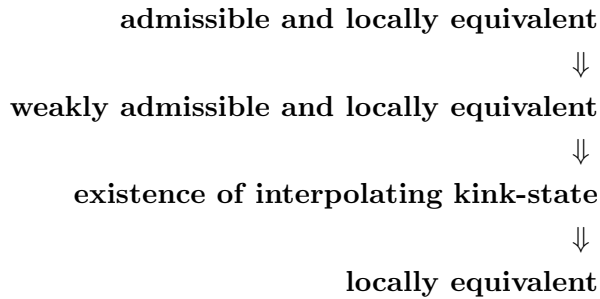
\begin{eqnarray*}
\mbox{\bf admissible and locally equivalent}      \\
\Downarrow                                   \\
\mbox{\bf weakly admissible and locally equivalent}\\ 
\Downarrow                                    \\
\mbox{\bf existence of interpolating kink-state} \\
\Downarrow                                      \\
\mbox{\bf locally equivalent}\\
\end{eqnarray*}

To motivate the following analysis, 
we give here the main steps of the construction of a kink-state.
A detailed discussion of our program is carried out in section 4 and 5.
Let as briefly summarize the strategy of it.
 
Firstly, we build the  
tensor-product of two copies of our QFT, i.e. the net
$\2O\mapsto \6A_2(\2O):=\6A(\2O)\otimes\6A(\2O)$ which we call in the sequel 
the {\em squared theory}. 
We consider the map $\al_F$, called the {\em flip automorphism},
which is given by interchanging the
tensor-factors, i.e.:
$$
\al_F:a_1\otimes a_2\mapsto a_2\otimes a_1
$$
The requirement 
that the pair of vacuum-states $\om_1,\om_2$ satisfies our technical 
assumptions guarantees the existence of an automorphism $\al_\te$
which has the following properties:  
\begin{description}
\item[{\it (1)}]
The automorphism $\al_\te$ is an involution, i.e. $\al_\te^2=\id$.
\item[{\it (2)}]
There exists a bounded region $\2O$, such that for each 
observable $a$ which is localized in the left space-like complement of
$\2O$ we have $\al_\te(a)=a$, and for each  
observable $a$ which is localized in the right space-like complement of
$\2O$ we have $\al_\te(a)=\al_F(a)$.
\end{description}
In the final step we introduce the algebra homomorphism
$$
\Delta_\te:a\in\6A\mapsto\al_\te(a\otimes\11)\in\6A\otimes\6A
$$
and show that the state 
$$
\om:=\om_1\otimes\om_2\circ\Delta_\te
$$
is a kink-state which interpolates $\om_1$ and $\om_2$. 
The interpolating property of $\om$ follows directly from its 
definition. The hard part is to prove that $\om$ satisfies the 
Borchers-criterion. We can also consider the kink-state
$$
\bar\om:=\om_2\otimes\om_1\circ\Delta_\te
$$
which is the anti-kink-state with respect to $\om$.
 
Let $x\mapsto U(x)=e^{iPx}$ be the 
representation of the translation group which corresponds to 
the kink-state $\om$ and $x\mapsto U_j(x)=e^{iP_jx}$ 
the representation of the translation group which corresponds to 
the vacuum-state $\om_j$, $j=1,2$.
If the vacuum states $\om_1$ and $\om_2$ are different, then 
$\om$ is not translationally invariant which implies that $0$ is not 
contained in the spectrum of the generator $P$. We show that the square of the
mass-operator $M^2=P_\mu P^\mu$ in the kink-representation
is bounded from below 
by $1/2\cdot\min (\inf(U_1),\inf(U_2))$, with 
$\inf(U_j):=\inf(\sp(P_{j,\mu}P_j^\mu)\backslash \{0\})$.
If there is a mass-gap in the theory, then we have $\inf(U_j)=\mu_j>0$
and obtain a lower bound estimate for the {\em kink-mass} $m$:
$$
m\geq 1/2 \min(\mu_1,\mu_2)
$$
These estimate is discussed in section 6.

We conclude this paper with section 7 "Conclusion and Outlook".
  
\section{Preliminaries}
Let us consider a quantum field theory in two dimensions which is described 
by a translationally covariant Haag-Kastler-net. We repeat the axioms of such a net
briefly.

A Haag-Kastler-net with translation covariance is a prescription that assigns
to each open double cone $\2O$ a C*-algebra $\6A(\2O)$ such that this 
prescription is isotonous and respects locality, i.e. $\2O_1\subset\2O_2$
implies $\6A(\2O_1)\subset\6A(\2O_2)$ and $\2O_1\subset\2O_2'$
implies $[\6A(\2O_1),\6A(\2O_2)]=\{0\}$. Here $\2O'$ denotes the space like
complement of a double cone $\2O$. Furthermore, let $\6A$ be the 
C*-inductive limit of the net $\2O\mapsto\6A(\2O)$ we assume the
existence of a group-homomorphism $\al:\7R^2\to\aut\6A$ from the translation
group into the automorphism group of $\6A$, such that $\al_x$ maps
$\6A(\2O)$ onto $\6A(\2O+x)$. 

We consider now a class of states of $\6A$ which are of interest for our 
sequel analysis. Let us select the set 
$\4S_A$ ($\4S_A^w$) of all {\em admissible (weakly admissible) vacuum states}
which consists of {\em pure vacuum states} $\om$ such that 
\begin{description}
\item[{\it (a)}] 
wedge duality holds as well as Haag duality in the GNS-representation, i.e.
$$
\6A_\om(W_\pm+x)=\6A_\om(W_\mp+x)' 
$$
and for $\2O=W_++x\cap W_-+y$ we have 
$$
\6A_\om(\2O)=\6A_\om(W_++x)\cap\6A_\om(W_-+y) \ \ \ .   
$$
\item[{\it (b)}] 
The GNS-representation $\pi$ of $\om$ is faithful.
\footnote{Condition {\it (c)} is only a proper assumption if $\6A$ is not 
simple.}
\item[{\it (c) Admissible:}]
for $W_\pm+x\subset W_\pm+y$ the inclusion
$$
\6A_\om(W_\pm+x)\subset\6A_\om(W_\pm+y)
$$
is a split inclusion or
\item[{\it (d) Weakly admissible:}]
there exists an automorphism $\beta$ of the C*-algebra
$$
\overline{\bigcup_{\2O} \6A_\om(\2O)\otimes \6A_\om(\2O)}^{||\cdot||}
$$
such that there is a space-like vector $r\in W_\pm$ and 
for each $a_1,a_2\in\6A_\om(\2O)$ holds the relation
\begin{displaymath}
\beta(a_1\otimes a_2)= \left\{
\begin{array}{ll} a_2\otimes a_1 & \mbox{if $\2O\subset W_\pm+r$} \\
                  a_1\otimes a_2 & \mbox{if $\2O\subset W_\mp-r$}
        \end{array}
      \right.
\end{displaymath}
and the automorphism
$$
\al_x\circ\beta\circ\al_{-x}\circ\beta
$$
is inner.
\end{description}

We now give a few comments on the notation used above. For a region  
$G\subset\7R^2$, $\6A(G)$ denotes the C*-algebra which is generated by all
$\6A(\2O)$'s with $\2O\subset G$. For a state $\om$ we write 
$\6A_\om(G)$ for the v.Neumann-algebra $\pi(\6A(G))''$, where 
$(\6H,\pi,\Om)$ is the GNS-triple of $\om$ and the double-prime denotes
the weak closure in $\6B(\6H)$. Moreover, $W_\pm$ denotes the wedge-region
$\{x| \ |x^0|< \pm x^1\}$.
It is clear that in two dimensions each double-cone $\2O$ 
is an intersection of two 
unique wedge-regions, i.e. $\2O=W_++x\cap W_-+y$.

{\em Remark:}
The condition {\em weakly admissible} looks rather technical and one
might worry that it can never be fulfilled. However, these condition 
can be proven for a large class of vacuum-states, namely the 
vacuum-states of the $P(\phi)_2$-models, whereas {\em admissibility}
can only be proven for the massive free scalar field \cite{Schl3,Schl4}.
Furthermore, we are going to see that each vacuum state 
which is admissible is also weakly admissible.      

\section{Definition of Kink-States}
In this section, we give a mathematical definition of kink-states and consider 
some immediate implications.

\bdef
A state $\om$ of $\6A$ is called a {\em kink-state} interpolating vacuum states
$\om_1,\om_2$ if 
\begin{description}
\item[{\it (a)}]
$\om$ satisfies the Borchers criterion.
\item[{\it (b)}]
Let $(\6H,\pi,\Om),(\6H_j,\pi_j,\Om_j)$ be the GNS-triples of
$\om,\om_j;j=1,2$.
$$
\pi|_{\6A(W_-)}\cong\pi_1|_{\6A(W_-)}
$$
$$
\pi|_{\6A(W_+)}\cong\pi_2|_{\6A(W_+)}    
$$
Here $\cong$ means {\em unitarily-equivalent}.
\end{description}
The set of all kink-states which interpolate $\om_1$ and $\om_2$ is denoted by
$\4S_{kink}(\om_1,\om_2)$.
\eef

{\em Remark:} In the sequel we write 
$(\6H,\pi,\Om),(\6H_j,\pi_j,\Om_j)$ for the GNS-triples of
$\om,\om_j;j=1,2$, unless we state something different.

Of course, since $\om$ is translationally covariant, we conclude for each 
$x\in\7R^2$:
\beq
\pi|_{\6A(W_-+x)}\cong\pi_1|_{\6A(W_-+x)}
\eeq
\beq
\pi|_{\6A(W_++x)}\cong\pi_2|_{\6A(W_++x)}    
\eeq
We show now that if there exists a kink-state $\om\in\4S_{kink}(\om_1,\om_2)$,
then the GNS-representations $\pi_1$ and $\pi_2$ 
are unitarily equivalent on each algebra $\6A(\2O)$.
We call two states which satisfies these relation {\em local 
unitarily equivalent} and write: $\om_1\cong_{loc}\om_2$.

\blem
If there exists a kink-state $\om\in\4S_{kink}(\om_1,\om_2)$ for  
$\om_1,\om_2\in\4S_A$, then $\om,\om_1,\om_2$ are 
local unitarily equivalent, i.e.:
$$
\4S_{kink}(\om_1,\om_2)\ni\om \ \ \ra \ \ 
\om \cong_{loc}\om_1\cong_{loc}\om_2
$$
\elem

\bpr 
As mentioned above, each double-cone can be written as 
$\2O=W_++x\cap W_-+y$. 
Now, let $\om\in\4S_{kink}(\om_1,\om_2)$ be a kink state.
We obtain from equ. (1) and (2): 
\beq
\pi|_{\6A(W_-+y)}\cong\pi_1|_{\6A(W_-+y)}
\eeq
\beq
\pi|_{\6A(W_++x)}\cong\pi_2|_{\6A(W_++x)}    
\eeq
Using Haag duality we obtain:
$$
\pi|_{\6A(\2O)}\cong\pi_1|_{\6A(\2O)}\cong\pi_2|_{\6A(\2O)}
$$
Since the double-cone $\2O$ can be chosen arbitrarily, the result follows.
\epr

{\em Remark:} Lemma 2.1 states that local unitary equivalence of two 
vacuum states $\om_1,\om_2$ is an essential condition for the existence of an
interpolating kink-state.


\section{The Existence of Kink-States}


In this section we formulate the main result of our paper which states that
for two weakly admissible vacuum states $\om_1,\om_2$ local unitary
equivalence is not only essential but also sufficient for the existence 
of kink-states. 

\bthe
For each pair of admissible or weakly admissible 
vacuum states $\om_1,\om_2\in\4S_A$ which are
local unitarily equivalent, i.e. $\om_1\cong_{loc}\om_2$, there exists
a kink-state $\om\in\4S_{kink}(\om_1,\om_2)$ which interpolates 
$\om_1$ and
$\om_2$.
\ethe

We do not give a proof of the theorem in this section 
because we need some further results for preparation. The complete 
proof is given in section 5.  
To motivate the sequel steps, we describe the construction of kink-states
briefly.
 
\begin{description}
\item[{\it (1)}]
The first step is to tensor two copies of our QFT, i.e. we consider the net
$\2O\mapsto \6A_2(\2O):=\6A(\2O)\otimes\6A(\2O)$. 
We introduce the 
map $\al_F$ which is given by prescription
$$
\al_F:a_1\otimes a_2\mapsto a_2\otimes a_1
$$
and extends to an automorphism of $\6A_2$ which is called
the {\em flip automorphism}.
\item[{\it (2)}]
Using the split property (in case of admissible vacuum-states), 
we can choose an unitary operator 
$\te\in\6A_{11}(W_++r)$ which implements the flip-automorphism
$\al_F$ on $\6A_2(W_+)$ ($W_+\subset W_++r$). 
Here and in the sequel we write 
$\6A_{ij}(G):=[ \ \pi_i\otimes\pi_j(\6A_2(G)) \ ]''$ 
for a region $G\in\7R^2$ and
$i,j=1,2$. 
\item[{\it (3)}]
We define the representation 
$\al_\theta:=\pi_{11}^{-1}\circ\Ad(\te)\circ\pi_{11}$ and show that 
$\al_\theta$ is an automorphism of an extension $\6F_2\supset\6A_2$
of $\6A_2$ and satisfies condition {\it (d)} of section 1.
In case of weakly admissible vacuum-states the existence of such an 
automorphism is required.  
\item[{\it (4)}]
In the final step we define the representation
$\rho:\6A\to\6B(\6H_1\otimes\6H_2)$ by
$$
\rho(a):=\pi_1\otimes\pi_2(\Delta_\te(a))
$$
with $\Delta_\te:=\al_\te|_{\6A\otimes\11}$.
We prove that $\rho$ is well defined and show that the 
state 
$$
\om:=\om_1\otimes\om_2\circ\Delta_\te
$$
is a kink-state which interpolates $\om_1$ and $\om_2$.
\end{description}


\section{Kink-States in the Squared Theory}


We consider now the tensor-product 
$\2O\mapsto\6A_2(\2O):=\6A(\2O)\otimes\6A(\2O)$ of the theory with itself
which is also called the squared theory. 
We show that for two local unitarily equivalent
admissible vacuum states $\om_1,\om_2$ of $\6A$ there is 
a canonical construction for a kink-state on the squared 
theory which interpolates the vacuum states $\om_1\otimes\om_2$
and $\om_2\otimes\om_1$.
 
So let us formulate the result of this section:

\bpro
For each pair of admissible vacuum states $\om_1,\om_2\in\4S_A$ which are
local unitarily equivalent, i.e. $\om_1\cong_{loc}\om_2$, there exists
a kink-state $\om\in\4S_{kink}(\om_{12},\om_{21})$ which interpolates 
$\om_{12}:=\om_1\otimes\om_2$ and
$\om_{21}:=\om_2\otimes\om_1$.
\epro

We start now with the preparation of the proof. Since the inclusion
$$
\6A_{11}(W_\pm+x)\subset\6N_{11}\subset\6A_{11}(W_\pm+y)
$$
is split with an intermediate type I factor $\6N_{11}$, the
flip-automorphism $\al_F$ is implemented on $\6A_2(W_+)$ 
in the representation $\pi_{11}:=\pi_1\otimes\pi_1$ by a 
unitary involution $\te$ ($\te^2=\11$) which is contained in 
$\6A_{11}(W_++r)$, i.e.
\beq
\pi_{11}\circ\al_F|_{\6A_2(W_+)}=\Ad(\te)\circ\pi_{11}|_{\6A_2(W_+)} \ \ \ . 
\eeq

For technical reasons, we introduce an extension of the net of 
local C*-algebras. For a vacuum state $\om\in\4S$, we define a family 
of semi norms on $\6A(\2O)$
\beq
||a||^\om_T:=|\tr(T\pi(a))|
\eeq
and denote by $\6M_\om(\2O)$ the closure of $\6A(\2O)$ in the topology which 
is induced by these family. It is clear that the algebra $\6M_\om(\2O)$ is a 
W*-algebra, canonical isomorphic to the v.Neumann algebra $\pi(\6A(\2O))''$.
The C*-inductive limit, generated by all $\6M_\om(\2O)$ is denoted by 
$\2A_\om$.

Consider now local unitarily equivalent vacuum states $\om_1,\om_2$ then 
we obtain 
$$
\6M(\2O):=\6M_{\om_1}(\2O)=\6M_{\om_2}(\2O) 
$$
and hence $\2A=\2A_{\om_1}=\2A_{\om_2}$.
There are unique extensions of the GNS-representations of $\om_1$ and $\om_2$
which are denoted by $\pi_1$ and $\pi_2$.

Let us consider now the net $\2O\mapsto\6F_2(\2O):=\6M(\2O)\otimes\6M(\2O)$.
We denote the C*-inductive limit which is generated by all 
algebras $\6F_2(\2O)$ by $\2A_2$ and
define the following representation of $\2A_2$:
\beq
\al^1_\te(a):= \te\pi_{11}(a)\te
\eeq

\blem
For each double cone $\2O$ which contains $\2O_r:=W_++r\cap W_-$ 
one has 
$$
\al^1_\te(\6F_2(\2O))\subset\6A_{11}(\2O)
$$
i.e. $\al^1_\te$ maps local algebras into local algebras.
\elem

\bpr
We choose an operator $b\in\6A_2(W_++y)$ such that for $W_+\supset W_++y$. 
Then we have:
$$
\begin{array}{l}
\te\pi_{11}(a)\te\pi_{11}(b)= \te\pi_{11}(a)\pi_{11}(\al_F b)\te
\vs
=\te\pi_{11}(a\al_F b)\te=\te\pi_{11}(\al_F b \  a)\te
\vs
=\pi_{11}(b)\te\pi_{11}(a)\te
\end{array}
$$
This implies that $\te\pi_{11}(a)\te$ is contained in 
$\6A_{11}(W_++y)'=\6A_{11}(W_-+y)$. On the other hand it is clear that
for $W_++x\subset W_+$ the operator $\te\pi_{11}(a)\te$ 
is also contained in $\6A_{11}(W_++x)$. Since each double cone
$\2O$ which contains $\2O_r$ can be written as $\2O=W_++y\cap W_-+x$, 
we conclude for each 
$a\in\6A_2(\2O)$:
\beq
\te\pi_{11}(a)\te\in\6A_{11}(W_-+y)\cap\6A_{11}(W_++x)=\6A_{11}(\2O) 
\eeq
which completes the proof.
\epr

Since $\pi_{11}$ is a faithful representation of $\2A$, it follows from
Lemma 3.1. that  
the prescription $\al_\te:a\mapsto \pi_{11}^{-1}(\al^1_\te(a))$ is a 
well defined endomorphism of $\2A$.

An immediate consequence of Lemma 3.1 is the following corollary:

\bcor
The automorphism $\al_\te\in\aut(\2A)$ has the following properties:
\begin{description}
\item[{\it (1):}]
$\al_\te^2=\id$ 
\item[{\it (2):}]
$\al_\te|_{\6F_2(W_+)}=\al_F|_{\6F_2(W_+)}$ and 
$\al_\te|_{\6F_2(W_-+r)}=\id_{\6F_2(W_-+r)}$
\end{description}
\ecor

We define now the following representation of $\6A_2$:
\beq
\rho:=(\pi_1\otimes\pi_2)\circ\al_\te
\eeq
Since $\al_\te$ is an automorphism, we obtain a further 
corollary:

\bcor
The representation $\rho$ is irreducible.
\ecor

We prove now that $\rho$ is a translationally covariant representation.

\blem
For each $x\in\7R^2$ is the automorphism 
$$
\al_{-x}\circ\al_\te\circ\al_x\circ\al_\te
$$
inner.
\elem

\bpr
Since $\te$ implements the flip-automorphism on $\6A_2(W_+)$ the operator 
$\te(x):=U_{11}(x)\te U_{11}(-x)$ implements $\al_F$ on $\6A_2(W_++x)$.
From this we obtain for $W_+\supset W_++x$ and 
$a\in\6A_2(W_++x)$:
\beq
\pi_{11}(\al_Fa)=\te\pi_{11}(a)\te=\te(x)\pi_{11}(a)\te(x)
\eeq
Hence $\te(x)\te$ is contained in $\6A_{11}(W_-+x)$.
On the other hand $\te(x)\te$ is contained in  $\6A_{11}(W_++r)$ with
$W_++r\supset W_+\supset W_++x$ and we obtain:
$$
\te(x)\te\in\6A_{11}(W_-+x)\cap\6A_{11}(W_++r)
$$
For the case $W_+\subset W_++x$ we obtain analogously:
$$
\te(x)\te\in\6A_{11}(W_-)\cap\6A_{11}(W_++r+x)
$$
By using Haag duality, we conclude that 
for each $x$ the operator $\te(x)\te$ is contained in $\6A_{11}(\2O_x)$,
where $\2O_x$ is a sufficiently large double cone. 
For each $x$ we define the unitary operator 
\beq
\gam(x):=\pi_{11}^{-1}(\te(x)\te)\in\6F_2(\2O_x)
\eeq
This implies the following relation
\beq
\begin{array}{l}
\al_x\circ\al_\te\circ\al_{-x}\circ\al_\te
\vs
=\pi_{11}^{-1}\circ\Ad(\te(x)\te)\circ\pi_{11}
\vs
=\Ad(\gam(x))
\end{array}
\eeq
which completes the proof.
\epr

\bcor
Each admissible vacuum-state is weakly admissible, i.e. 
$$
\5S_A\subset \5S_A^w \ \ .
$$
\ecor
\bpr
By Corollary 4.1 and
Lemma 4.2, the automorphism $\beta=\al_\te$ satisfies condition {\it (d)}
of section 1.
\epr

In the sequel analysis, it is sufficient to consider vacuum-states which are 
{\em weakly admissible}.  

\bcor
The representation $\rho$ is translationally covariant, i.e. 
there exists a strongly continuous representation $x\mapsto U_\rho(x)$
of the translation group which implements $\al_x$:
$$
\rho\circ\al_x = \Ad(U_\rho(x))\circ\rho
$$
\ecor

\bpr
In the sequel, we use the following abbreviations:
$$
U_{ij}(x):=U_i(x)\otimes U_j(x)
$$
where $U_i$ implements the translation group in the vacuum representation 
$\pi_i$; $i,j=1,2$. 
By construction, $\gamma(x)$ satisfies the following {\em co-cycle condition}:
\beq
\gam(x+y)=\al_x(\gam(y))\gam(x)
\eeq
Moreover, the following relation holds
\beq
\al_\te\circ\al_x=\al_x\circ\al_{\te(-x)}
\eeq
and we obtain by using Lemma 4.2
\beq
\begin{array}{l}
\rho\circ\al_x=\pi_{12}\circ\al_\te\circ\al_x
\vs
=\pi_{12}\circ\al_x\circ\al_{\te(-x)}
\vs
=\Ad(U_{12}(x))\circ\pi_{12}\circ\al_{\te(-x)}
\vs
=\Ad(U_{12}(x))\circ\pi_{12}\circ\Ad(\gam(-x))\circ\al_\te
\vs
=\Ad(U_{12}(x)\Gam_\rho(-x))\circ\rho
\end{array}
\eeq
where we have set $\Gam_\rho(x):=\pi_{12}(\gam(x))$.
Hence for each $x\in\7R^2$ we obtain the charge transporter 
\beq
\Gamma_\rho(x):=\pi_{12}(\gam(x))\ \ \ .
\eeq
We define 
$$
U_\rho:x\mapsto U_\rho(x):=U_{12}(x)\Gamma_\rho(-x)
$$ 
which is a strongly continuous representation of the translation group
implementing the translations in the representation $\rho$.
\epr

To study the properties of the representation $\rho$ in more detail 
we consider a further representation $\bar\rho:\6A_2\to\6B(\6H_2\otimes\6H_1)$
which is given by 
\beq
\bar\rho:=\pi_{21}\circ \al_\te  \ \ \ .
\eeq
We will see that $\bar\rho$ plays 
the role of an {\it anti-kink}.
The representation $\bar\rho:\6A_2\to\6B(\6H_2\otimes\6H_1)$ is 
translationally covariant which can be proven by using 
the arguments in the proofs of 
Lemma 4.1 and Lemma 4.2 for $\bar\rho$.

To establish the statement of Proposition 4.1, it remains to be proven, 
that the spectrum of $U_\rho$ is contained in the closed forward 
light cone. For this purpose, we prove now the additivity of 
energy momentum spectrum. 

Let us consider two representations 
$\rho_1=\pi_{12}\circ\al_{\te_1}$
and $\rho_2=\pi_{21}\circ\al_{\te_2}$, where $\te_1$ resp. $\te_2$ are 
unitary operators which implement $\al_F$ on $\6F_2(W_++x_1)$ resp.
$\6F_2(W_++x_2)$, or in case of weakly admissible vacuum-states
$\al_{\te_1}$ and $\al_{\te_2}$ are automorphisms which satisfy
condition {\it (d)} of section 1. 
Then we define the composition of 
$\rho_1$ and $\rho_2$ as follows:
\beq
\rho_1\rho_2:=\rho_1\circ\pi_{21}^{-1}\circ\rho_2=
\pi_{12}\circ\al_{\te_1}\al_{\te_2}
\eeq
The representations can also be composed in the other direction:
\beq
\rho_2\rho_1:=\pi_{21}\circ\al_{\te_2}\al_{\te_1}
\eeq
Moreover, we write in the sequel $S(\rho)$ for the spectrum of $U_\rho$.

{\it Remark:} The composition described above can be interpreted 
as the composition of soliton homomorphisms in the sense of \cite{Schl2}.
Remember that $\rho\circ\pi_{21}^{-1}$ maps $\6A_{21}$ into $\6A_{12}$.

\blem 
Let $\rho_1,\rho_2$ and $\rho_1\rho_2$ be defined as described above,
then the additivity of the energy-momentum spectrum holds, i.e.
$$
S(\rho_1)+S(\rho_2)\subset S(\rho_1\rho_2)
$$
\elem

\bpr
The proof is standard and uses the same method as in the DHR-framework
\cite{DHRI,DHRII}.
The only difference which appears is due to the fact that 
the representations $\rho_j$ are localized in wedge regions and not in 
double cones. But for the proof it is sufficient that $\rho_j$ maps 
local algebras into local algebras. 

We choose test 
functions $f_j$ with $\supp \tl f_j\subset S(\rho_j)$ and a local operator
$a\in\6F_2(\2O)$. The operators
$$
a_j:=\int dx \ f_j(x) \gam(x) \al_x a
$$
have energy-momentum transfer in $\supp \tl f_j$. Here  
$\gam(x)$ is defined as in equ.(11) above.
Now $\Psi_1=\pi_{12}(a_1)\Om_{12}\in\6H_1\otimes\6H_2$ has 
energy-momentum support in $\supp \tl f_1$ and 
$\Psi_2=\pi_{21}(a_2)\Om_{21}\in\6H_2\otimes\6H_1$ has 
energy-momentum support in $\supp \tl f_2$.
Moreover, the vector 
$$
\Psi:=\rho_1(a_2)\pi_{12}(a_1)\Om_{12}
$$
has energy-momentum support in $\supp \tl f_1 +\supp \tl f_2$ 
which remains also true for 
$$
\Psi_y:=\rho_1(a_2)U_{\rho_1}(y)\pi_{12}(a_1)\Om_{12} \ \ . 
$$
We compute now:
$$
||\Psi_y||^2
=\<\Om_{12},\pi_{12}(a_1^*)\rho_1(\al_{-y}(a_2^*a_2))\pi_{12}(a_1)\Om_{12}\>
$$
Since $\rho_1$ acts as $\pi_{12}\circ\al_F$ on $\6F_2(W_++y)$ with 
$W_++y\subset W_++x_1$, we conclude by using the cluster theorem:
$$
\begin{array}{l}
\lim_y ||\Psi_y||^2
=||\Psi_1||^2 \
\<\Om_{12},\pi_{12}(\al_F(a_2^*a_2))\Om_{12}\>
\vs
=||\Psi_1||^2 \
\<\Om_{21},\pi_{21}(a_2^*a_2)\Om_{21}\>
\vs
=||\Psi_1||^2 \
||\Psi_2||^2
\end{array}
$$
as $y$ tends to minus space like infinity.
Hence for $||\Psi_j||\not= 0$ we obtain $\Psi_y\not= 0$ for 
one $y\in\7R^2$ and the result follows.
\epr

Let us have a closer look at the anti-kink representation $\bar\rho$. 
We denote by $J_{kl}$ the modular conjugation with respect to the pair
$(\6A_{kl}(W_+),\Om_{kl})$. For technical reasons, we make the following
assumption:

{\it Assumption:} Let us assume that there exists a PCT-symmetry, i.e. 
an involutive anti-automorphism 
$j:\6A_2\mapsto\6A_2$ with $j(\6A_2(\2O))=\6A_2(-\2O)$ 
and $j\circ\alpha_x=\alpha_{-x}\circ j$ which is 
implemented in each vacuum representation  $\pi_{kl}$ by the
modular conjugation $J_{kl}$, i.e.:
$$
\pi_{kl}(ja)=J_{kl}\pi_{kl}(a)J_{kl}
$$

Now we define the following representation:
\beq
\rho^J:=j_{21}\circ\rho_F\circ j
\eeq
Here we have set $j_{kl}:=\Ad(J_{kl})$ and the representation $\rho_F$ is given
by
\beq
\rho_F:=\pi_{21}\circ\al_F\circ\al_\te \ \ \ .
\eeq

{\it Remark:}
If an automorphism $\beta$ of $\2A_2$ satisfies the condition {\it (d)}
of section 1, then the
automorphism $\al_F\circ\beta$ satisfies 
it also.

\blem
The representations $\bar\rho$ and $\rho^J$ are unitarily equivalent and in 
addition we obtain that $S(\rho)=S(\bar\rho)$.
\elem

\bpr
Using the composition role described above, we obtain that 
$\rho\bar\rho=\pi_{12}$ and $\bar\rho\rho=\pi_{21}$. 
By Corollary 4.2 and Lemma 4.2 we can use the results of
\cite{GuiLo,Lo2,Schl2} 
we conclude that the anti-kink sector is unique. Thus we have
$\bar\rho\cong\rho^J$.
In addition to that, the representations $\rho$ and $\rho_F$ 
are unitarily equivalent
($\rho_F$ is a PT-conjugate for $\rho$ in the sense of \cite{Schl2}).
Since $U_{\rho^J}(x):=J_{21}U_{\rho_F}(-x)J_{21}$ implements the translation
group in the representation $\rho^J$ (see also \cite{Bor1,GuiLo,Schl2}), 
we conclude
$$
S(\bar\rho)=S(\rho^J)=S(\rho_F)=S(\rho)
$$
which completes the proof.
\epr

\bpro
$\rho$ is a positive energy representation.
\epro

\bpr
Corollary 4.2 and Lemma 4.2 state that $\rho$ is translationally covariant 
and irreducible, in particular factorial. Now by Lemma 4.3 
we conclude that 
$S(\rho)+S(\bar\rho)\subset S(\pi_{12})=S(\pi_{21})=S(\pi_1)+S(\pi_2)$
and with Lemma 4.4 we obtain finally 
\beq
S(\rho)\subset \bar V_+ 
\eeq
which completes the proof.
\epr

We are now ready to prove Proposition 4.1.
\ska\noindent
{\it Proof of Proposition 4.1:}
We show that $\om_\te:=\om_1\otimes\om_2\circ\al_\te$ is a kink-state 
which interpolates the vacuum states $\om_1\otimes\om_2$ and 
$\om_2\otimes\om_1$. Since $\rho$ is irreducible, 
the GNS-representation $\pi_\te$ of $\om_\te$ is unitarily equivalent to
$\rho$. Hence $\om_\te$ satisfies the Borchers criterion by 
Proposition 4.2. Furthermore, by Corollary 4.1 we conclude that 
$\om_\te$  interpolates the vacuum states $\om_1\otimes\om_2$, namely
we have
$$
\begin{array}{l}
\rho|_{\6A_2(W_+)}=\pi_{12}\circ\al_F\cong\pi_{21}
\vs
\rho|_{\6A_2(W_-+r)}=\pi_{12} 
\end{array}
$$
and the result follows. 
\epr


\section{Kink-States in the Original Theory}


We have seen in the last section that each QFT which is equipped with two 
different (admissible) vacuum states there is a method to
construct kink-states in the squared theory. We are now interested in 
the existence of kink-states for the original theory. 

Since we are able to construct kink-states for the squared theory it is easy 
to obtain kink-states for the original one. Let us consider the 
automorphism $\al_\te\in\aut(\2A_2)$ with was constructed in the last section
and define the following algebra-homomorphisms:
$$
\begin{array}{l}  
\Delta_\te:\2A\to\2A\otimes\2A \ \ \ ; 
\ \ \ a\mapsto \Delta_\te(a):=\al_\te(a\otimes\11)
\vs
\Delta'_\te:\2A\to\2A\otimes\2A \ \ \ ; 
\ \ \ a\mapsto \Delta'_\te(a):=\al_\te(\11\otimes a)
\end{array}  
$$
We obtain now states 
$$
\begin{array}{l}
\om_\te:=\om_1\otimes\om_2\circ\Delta_\te
\vs
\om'_\te:=\om_1\otimes\om_2\circ\Delta'_\te
\end{array}  
$$
which has the following localization properties:
\beq
\begin{array}{l}
\om_\te|_{\6A(W_+)}=\om_2|_{\6A(W_+)} 
\ \ \ \om_\te|_{\6A(W_-+r)}=\om_1|_{\6A(W_-+r)}
\vs
\om'_\te|_{\6A(W_+)}=\om_1|_{\6A(W_+)}
\ \ \ \om'_\te|_{\6A(W_-+r)}=\om_2|_{\6A(W_-+r)}
\end{array}  
\eeq
We use now the results of the last section to prove that both 
$\om_\te$ and $\om_\te'$ are kink states which also proves Theorem 3.1.

\bpro
The states $\om_\te$ and $\om_\te'$ are kink-states where 
$\om_\te$ is contained in $\4S_{kink}(\om_1,\om_2)$ and $\om'_\te$ 
is contained in $\4S_{kink}(\om_2,\om_1)$.
Moreover, each state $\hat\om$ which GNS-representation is a sub-representation
of the GNS-representation of $\om_\te$ is also contained in 
$\4S_{kink}(\om_2,\om_1)$.
\epro

\bpr
By construction, $\om_\te$ is the restriction of the state 
$\om_1\otimes\om_2\circ\al_\te$ to the first tensor factor, i.e. the 
algebra $\2A\otimes\7C\11$. We show now that the  
GNS-representation $\sgm$ of $\om_\te$ is unitarily equivalent to 
$\rho|_{\2A\otimes\7C\11}$.

The C*-algebra $\6A(W_-+r,W_+)$ which is 
generated by $\6A(W_-+r)$ and $\6A(W_+)$ is contained in $\6A$.
By using the Theorem of Reeh and Schlieder, we obtain that
$$
\rho(\6A(W_-+r,W_+))\otimes \11)\Om_1\otimes\Om_2=
\pi_1(\6A(W_-+r))\Om_1\otimes\pi_2(\6A(W_+))\Om_2
$$
is dense in $\6H_1\otimes\6H_2$. Hence 
the representation $\rho|_{\2A\otimes\7C\11}$ 
is cyclic and therefore unitarily equivalent to $\sgm$.

Since $\rho$ is a positive energy representation (Proposition 4.2) of
$\2A\otimes\2A$ its restriction $\rho|_{\2A\otimes\11}\cong\sgm$ is a 
positive energy representation of $\2A$. By a result of Borchers 
\cite{Bor2}, we can construct a unitary 
strongly continuous representation $x\mapsto U_\sgm(x)$ 
of the translation
group with the following properties:
\begin{description}
\item[{\it 1:}]
For each $x$ is the operator $U_\sgm(x)$ contained in $\sgm(\2A)''$. 
\item[{\it 2:}]
$U_\sgm$ implements the translations in the representation $\sgm$, 
i.e. $\sgm(\al_xa)=U_\sgm(x)\sgm(a)U_\sgm(-x)$.
\item[{\it 3:}]
The spectrum of $U_\sgm$ is contained in the closed forward light cone.
\end{description}
Now let $(\pi,\6H)$ be a sub-representation of $\sgm$, i.e. there exists an 
isometry $v:\6H\to\6H_1\otimes\6H_2$ such that 
$\pi=v^*\sgm(\cdot )v$. Since $vv^*$ is a projection which is contained 
in $\sgm(\2A)'$ we conclude that $U_\pi(x):=v^*U_\sgm(x)v$ is a unitary 
strongly continuous representation of the translations which
implements the translations in the representation $\pi$. In particular 
the spectrum of $U_\pi$ is also contained in the closed forward light cone.
Thus $\pi$ satisfies the Borchers criterion.

From equ. (22) we obtain the following relations: 
\begin{eqnarray*}
\sgm|_{\6A(W_-+r)}
\cong\pi_1|_{\6A(W_-+r)}\otimes\11
\cong_{quasi}\pi_1|_{\6A(W_-+r)}
\vs\vs
\sgm|_{\6A(W_+)}
\cong\11\otimes\pi_2|_{\6A(W_+)}
\cong_{quasi}\pi_2|_{\6A(W_+)}
\end{eqnarray*}
Here the symbol $\cong_{quasi}$ means {\em quasi-equivalent}.
Since $\pi$ is a sub-representation of $\sgm$, we conclude: 
\begin{eqnarray*}
\pi|_{\6A(W_-+r)}\cong_{quasi}\pi_1|_{\6A(W_-+r)}
\vs\vs
\pi|_{\6A(W_+)}\cong_{quasi}\pi_2|_{\6A(W_+)}
\end{eqnarray*}
Using the fact that the v.Neumann-algebras $\pi_1(\6A(W_-+r))''$ and 
$\pi_2(\6A(W_+))''$ are type III factors, we conclude by using 
standard-arguments:
\begin{eqnarray*}
\pi|_{\6A(W_-+r)}\cong\pi_1|_{\6A(W_-+r)}
\vs\vs
\pi|_{\6A(W_+)}\cong\pi_2|_{\6A(W_+)}
\end{eqnarray*} 

Thus the state $\om_\te$ is a kink-state and 
each state $\hat\om$ 
which GNS-representation is a sub-representation
of the GNS-representation of $\om_\te$ is also contained in 
$\4S_{kink}(\om_2,\om_1)$.
The proof for $\om_\te'$ works analogously.
\epr


\section{Estimates for the Kink- (Soliton-) Mass}


To discuss the mass of kink- (soliton-) state, we consider purely massive 
theories where the admissible vacuum states are {\it massive vacuum states}.

Let $\om$ be a pure translationally covariant state and 
$U:x\to U(x)$ the strongly continuous representation of
the translations which implements $\al_x$ in the GNS-representation
of $\om$. Then $\om$ is called a {\it massive vacuum state} if the spectrum 
of $U(x)$ contains $\{0\}$ and a subset of 
$C_\mu:=\{p\in\7R^2: p^2>\mu\}$ where $\mu>0$ is a positive real number,
called the {\em mass gap} of $\om$. 
We denote the set of all massive weakly admissible vacuum states 
with mass gap $\mu$ by $\5S(\mu)$.
If the spectrum of $U(x)$ contains the mass shell 
$H_m:=\{p\in\7R^2:p^2=m^2\}$ and a subset of $C_{\mu+m}$, then we call
$\om$ a massive one-particle state with mass $m>0$. 

For a two dimensional QFT it is shown \cite{BuFre,Fre2,Schl1}, that 
for each massive one-particle state $\om$ there are massive vacuum 
states $\om_1,\om_2$, such that $\om$ interpolates $\om_1$ and $\om_2$.
The mass $m$ of $\om$ then satisfies the estimate
\beq
m\geq {1\over 2}\min(\mu_1,\mu_2)
\eeq
where $\mu_1$ (resp. $\mu_2$) is the mass gap of $\om_1$ (resp. $\om_2$).

Now we consider the situation where two different massive admissible 
vacuum states $\om_1\in \5S(\mu_1)$ and
$\om_2\in \5S(\mu_2)$ are given. Then we know by Theorem 3.1 that 
there exist a kink-state $\om$ which interpolates $\om_1$ and $\om_2$.

We denote by $S(\pi)$ the spectrum of $U_\pi(x)$, where $U_\pi$ is a
strongly continuous representation of the translation group which implements
$\al_x$ in the GNS-representation $\pi$ of $\om$. 
 
If the vacuum states are $\om_1$ and $\om_2$ are inequivalent, then 
it follows that $0\notin S(\pi)$. This can be seen as follows:
Since $\om_1$ and $\om_2$ are inequivalent, there exists an 
operator $a\in\2A$ with $\om_1(a)\not=\om_2(a)$.
On the other hand, if $x$ tends to space-like infinity we have
$\lim_{|x|\to \infty}\om(\al_x a)=\om_2(a)$ and if 
$x$ tends to minus space-like infinity we have
$\lim_{|x|\to -\infty}\om(\al_x a)=\om_1(a)$ and $\om$ is not 
translationally invariant. 

From the proofs of Proposition 4.2 and Proposition 5.1 we obtain that
$S(\pi)$ is a subset of the closed forward light cone which 
does not contain the point $k=0$. Hence we conclude
\beq
S(\pi)\subset {1\over 2}(S(\pi_1)+S(\pi_2))
\eeq
and obtain for the infimum  
$\inf(S(\pi))$ of the spectrum $S(\pi)$ the estimate:
\beq
\inf(S(\pi))\geq {1\over 2}\min(\mu_1,\mu_2)
\eeq
Here the infimum $\inf(S(\pi))$ is defined as the 
the infimum of the spectrum of the mass operator 
$M=(P_\mu P^\mu)^{1/2}$, where $P$ is the generator of the translation 
group $U_\pi$. 

Let us suppose that $\om$ dominates a massive 
one particle state with mass $m>0$,
then we obtain from equ. (25) that $m$ satisfies the estimate 
of equ. (23), namely $m\geq 1/2\min(\mu_1,\mu_2)$.
 
We conclude this section by summarizing the discussion above.
If we consider a massive one-particle state $\om_m$ with mass $m>0$, 
then there are massive vacuum states $\om_1$ and $\om_2$ with corresponding
mass gaps $\mu_1,\mu_2$, such that $\om_m$ interpolates $\om_1$ and $\om_2$
and $m$ satisfies the estimate 
$m\geq 1/2\min(\mu_1,\mu_2)$. 
Using the result of section 3 (Theorem 3),
we can construct from the vacuum states $\om_1,\om_2$ 
a kink-state $\om$ which also interpolates $\om_1$ and $\om_2$. 
If there exists a purification $\om_{m'}$ of $\om$ which is 
a massive on particle state with mass $m'$, then $m'$ satisfies the 
same estimate as the mass $m$, namely 
$m'\geq 1/2\min(\mu_1,\mu_2)$. Here we have assumed that $\om_1$ and
$\om_2$ are admissible vacuum states.

\section{Conclusion and Outlook}
We have seen that for each pair of admissible vacuum states which are
also locally equivalent there is a natural way to construct 
an interpolating kink-state. One advantage of this construction is, 
that we do not need
the assumption that the vacua are related by an internal 
symmetry transformation as in \cite{Froh1}. Furthermore, the construction is 
purely algebraic and independent of the specific properties of a model.

On the other hand, if we want to apply our result to a concrete model, we
have to check that the vacuum states of 
the model of consideration are admissible or
weakly admissible. At the moment, admissibility is 
only checked for the massive free scalar field 
\cite{AntFre,Bu1,Schl3}. For the vacuum states of the 
$P(\phi)_2$-models  
one can prove weak admissibility \cite{Schl4}.

If we consider a massive one-particle state $\om_m$ with mass $m>0$, 
then there are massive vacuum states $\om_1$ and $\om_2$ with corresponding
mass gaps $\mu_1,\mu_2$, such that $\om_m$ interpolates $\om_1$ and $\om_2$.
If $\om_1$ and $\om_2$ weakly admissible vacuum-states,
then we can apply Theorem 3.1 and construct 
a kink-state $\om$ which also interpolates $\om_1$ and $\om_2$. 
It is not well understood at the moment, what are the relations 
between $\om_m$ and the reconstructed kink-state $\om$.    

\subsubsection*{{\it Acknowledgment:}}
I am grateful to Prof. K. Fredenhagen for 
supporting this investigation with many ideas. I am also grateful to 
Prof. C. d'Antoni for hints and discussion.
Thanks are also 
due to my colleagues in Hamburg for careful reading.



\begin{thebibliography}{References}
\bibitem{AntFre} d'Antoni, C. and Fredenhagen, K.:
Carges in Spacelike Cones, Commun. Math. Phys. {\bf 94}, 537-544 (1984)
\bibitem{AntLo} d'Antoni, C. and Longo, R.: Interpolation by Type I Factors 
and the Flip Automorphism, Jour. Func. Anal. {\bf 51}, 361-371 (1983) 
\bibitem{ArkHa}  Araki, H. and Haag, R.:
{Collision Cross Sections in Terms of
Local Observables.}, Commun. Math. Phys. {\bf 4}, 77-91, (1967)
\bibitem{Bor1} Borchers, H.-J.: {CPT-Theorem in the Theory of Local 
Observables}, Commun. Math. Phys. {\bf 143}, 315-332, (1992) 
\bibitem{Bor2} Borchers, H.-J.: {On the Converse of the 
Reeh-Schlieder Theorem}, Commun. Math. Phys. {\bf 10}, 269-273, (1968)  
\bibitem{Bor3} Borchers, H.-J.: Commun. Math. Phys. {\bf 4}, 315-323, (1967)
\bibitem{Bu1} Buchholz, D.:
{Product States for Local Algebras.}
Commun. Math. Phys. {\bf 36}, 287-304, (1974)
\bibitem{BuFre}  Buchholz, D.  and Fredenhagen K.:
{Locality and the Structure of Particle States.}
Commun. Math. Phys. {\bf 84}, 1-54, (1982)
\bibitem{DHRI}
Doplicher, S., Haag, R. and Roberts, J.E.: {Local Observables and
Particle Statistics I.} 
Commun. Math. Phys. {\bf 23}, 199-230, (1971)
\bibitem{DHRII}
Doplicher, S., Haag, R. and Roberts, J.E.: {Local Observables and
Particle Statistics II.} 
Commun. Math. Phys. {\bf 35}, 49-58, (1971)
\bibitem{Fre1}
Fredenhagen, K.:
{Generalization of the Theory of Superselection Sectors.}
Published in Kastler, D.:
{The Algebraic Theory of Superselection Sectors ...}
World Scientific 1989
\bibitem{Fre2}
Fredenhagen, K.: {Superselection Sectors in Low Dimensional
Quantum Field Theory.} , DESY-92-133, Sept 1992. 
Published in the proceedings of 28th Karpacz Winter School of
Theoretical Physics: Infinite-Dimensional Geometry in Physics,
Karpacz, Poland, 17-29 Feb 1992.
J. Geom. Phys. {\bf 11} (1993) 337-348  
\bibitem{Fre3}
Fredenhagen, K.: {On the Existence of Antiparticles.} 
Commun. Math. Phys. {\bf 79}, 141-151, (1981)
\bibitem{Froh1}
Fr\"ohlich, J.: {New Superselection Sectors (Soliton States)
in Two Dimensional Bose Quantum Field Models.}
Commun. Math. Phys. {\bf 47}, 269-310, (1976)
\bibitem{Froh3}
Fr\"ohlich, J.: {Quantum Theory of
Nonlinear Invariant Wave (Field) Equations.} Erice, Sicily, Summer 1977
\bibitem{GuiLo} 
Guido, D. and Longo, R.:
{Relativistic Invariance and Charge Conjugation in
Quantum Field Theory.}
Commun. Math. Phys. {\bf 148}, 521-551, (1992)
\bibitem{Ha} 
Haag, R.: {Local Quantum Physics.}
Berlin, Heidelberg, New York: Springer 1992
\bibitem{Lo1}
Longo, R.: {Index of Subfactors and Statistics I.}
Commun. Math. Phys. {\bf 126}, 217-247, (1989)
\bibitem{Lo2}
Longo, R.: {Index of Subfactors and Statistics II.}
Commun. Math. Phys. {\bf 130}, 285-309, (1990)
\bibitem{Schl1}
Schlingemann, D.: {Antisolitonen und Mehrsolitonzust\"ande
im Rahmen der Algebraischen Quantenfeldtheorie.} Diploma thesis,
Hamburg 1994
\bibitem{Schl2}
Schlingemann, D.: On the Algebraic Theory of Soliton and Anti-soliton
Sectors. DESY-95-012, Feb 1995. 23pp. to appear in Rev. Math. Phys.
\bibitem{Schl4}
Schlingemann, D.: 
Kink-States in $P(\phi)_2$-Models.
(In preparation) 
\bibitem{Schl3}
Schlingemann, D.: 
Remarks on the Split Property for the Massive Free Scalar-Field. 
(In preparation)
\end{thebibliography}
\end{document}